# ECRH beam broadening in the edge turbulent plasma of fusion machines.

E.V. Sysoeva[1, 2], F. da Silva[3], E.Z. Gusakov[1, 2], S. Heuraux[4], A.Yu. Popov[1, 2]

[1] *Ioffe Institute, St.Petersburg, Russia*

[2] *RL PAT SPbSPU, St.Petersburg, Russia*

[3] *Institute of Plasmas and Nuclear Fusion, IST.*

[4] *IJL (CNRS UMR-7198), Universite Henri Poincaré, 54506 Vandoeuvre Cedex, France*

## Introduction

Quasi-optical microwave beams possessing sufficiently small divergence are often used in fusion plasmas for the purpose of diagnostics (density profile reflectometers [1], Doppler reflectometry [2] and collective Thomson scattering (CTS) [3]) or electron cyclotron resonance heating and current drive (ECRH&ECCD) [4]. The O-mode beams are planned for application in ITER for heating [5] and tearing-mode stabilization [6] whereas the X-mode microwave beams are proposed for diagnostics [7]. Small angular width of these beams is essential for diagnostics wavenumber resolution whereas a modest change of the spatial beam width could be critical for the local ECRH power deposition and stabilization of tearing-mode. Turbulent plasma edge of magnetic fusion devices that changes with the different scenarios using different heating systems and plasma parameters can influence both angular and spatial width of the microwave beam leading to degradation of diagnostics resolution or broadening of the power deposition profile. This topic, which is comparatively hot, was studied in the framework of the ray tracing approach both numerically [8] and analytically in 2D geometry [9] assuming a filamentary structure of edge plasma turbulence and in 3D geometry [10] supposing spherically symmetric plasma density perturbations with Gaussian radial distribution. In both analytical approaches treating the ray diffusion the background plasma was supposed to be homogeneous and only the angular beam width was estimated.

The aim of the present paper is to derive simple analytical expressions for both angular and spatial width of the microwave beam applicable for future ITER experimental situation of the O-mode propagating through inhomogeneous edge plasma layer, where the turbulent density fluctuations elongated along the magnetic field are excited.

Two analytical models are developed here, the first one is based on the eikonal perturbation method introduced in statistical radio physics in the framework of WKB approach [11], and the second one uses the weak turbulent theory [12].

The analytical expressions for the microwave beam mean waist growth along the trajectory obtained using the eikonal perturbation method and depending on the turbulence and plasma parameters is compared to time-averaged results coming from 2D Maxwell's equations solver [13] in the case of ordinary mode propagation in turbulent inhomogeneous plasma for different turbulence k-spectra and plasma conditions going from model cases for validation to experimental and future situations to predict the behavior of the considered experiments.

The paper is organized as follows: in the first section explicit expressions for angular and spatial beam width are obtained with the help of the eikonal perturbation method. In the second section the angular broadening law is derived with the weak turbulence theory method. As a conclusion of the theoretical part the identity of equations for beam angular width obtained using both techniques is demonstrated. In the last section the comparison of analytical and numerical results are presented. To conclude we discuss predicted broadening of microwave beams for future ECRH experiments at ITER.

## 1. Beam spatial and angular broadening in the frame of Eikonal perturbation method approach

The eikonal perturbation method we are going to generalize to the case of inhomogeneous plasma for any density fluctuations distribution in the illuminated zone came from physics of turbulent atmosphere [11]. It was recently applied to describe the angular broadening of the probing beam in Doppler reflectometry [14]. As we develop theoretical models of the plasma edge turbulence impact on the microwave beam applicable for big ITER-size plasma, the curvature effects can be assumed negligible for this kind of study focused on the effects of the density fluctuations determined by the ITER plasma scenarios on beam properties.

Thus, we will consider the O-mode propagation in the slab background plasma inhomogeneous in a direction perpendicular to the external magnetic field. Let us choose the axes in the following way: axis z is directed along the external magnetic field and axis x is directed along the density gradient where x=0 stands for the plasma edge position. Under these approximations we can describe the O-mode electric field using the Helmholtz equation

$$\left[\frac{\partial^2}{\partial x^2}+\frac{\partial^2}{\partial y^2}+k(x)^2+\delta k(x,y)^2\right]E(x,y)=0 \quad (1)$$

where $\delta k^2=\frac{\omega^2}{c^2}\frac{\delta n(x,y)}{n_c}$, $k^2=\frac{\omega^2}{c^2}\left[1-\frac{\omega_{pe}^2}{\omega^2}\right]$

Supposing the turbulence correlation length larger than the wavelength we describe the wave propagation using the WKB approximation and, according to [11], with the accuracy to the second order in the density perturbation $\delta n(x,y)$ assume that it influences only the phase of the partial wave propagating at a given angle not affecting its amplitude spatial evolution and the ray trajectory. Under this assumption in paraxial approximation we get the microwave spatial structure in the form of partial waves superposition

$$E(x,y)\approx\int\exp\left\{i\left[\int_0^x k(x')dx'-\frac{k_y{}^2d^2(x)}{2}+k_y y-\frac{\omega^2}{c^2}\int_0^x\frac{\delta n\left(x',y'\left(y,x,x',k_y\right)\right)dx'}{2n_c k(x')}\right]\right\}\frac{A(k_y)dk_y}{2\pi\sqrt{k(x)}} \quad (2)$$

where

$$A(k_y)=\sqrt{k(0)}\int E(0,y)e^{-ik_y y}dy$$

stands for the antenna diagram determined by the field distribution at the plasma boundary,

$$d^2(x) = \int_0^x \frac{dx'}{k(x')},$$

$y'(y, x, x', k_y)$ is the ray trajectory possessing poloidal wavenumber $k_y$ and connecting points $(x', y')$ and $(x, y)$. In paraxial approximation we get

$$y'(y, x, x', k_y) = y - k_y \int_{x'}^{x} \frac{dt}{k(t)} \triangleq y - k_y l^2(x, x').$$

Each component of the antenna diagram (partial wave) suffers from the random phase modulation due to propagation in the turbulent plasma, resulting in the net microwave beam distortions which are described by expression (2). The field distribution provided by (2) is random, nevertheless the microwave beam angular width is related to its poloidal structure. To characterize the average angular and spatial structure of the field across the direction of propagation we use the two-point Cross-Correlation Function (CCF)

$$\left\langle E(x, y_1) E^*(x, y_2) \right\rangle = \iint \frac{dk_{y1} dk_{y2}}{(2\pi)^2 k(x)} A(k_{y1}) A(k_{y2}) e^{ik_{y1}y_1} e^{-ik_{y2}y_2} e^{-\frac{i}{2}k_{y1}^2 d^2(x)} e^{\frac{i}{2}k_{y2}^2 d^2(x)} \cdot \left\langle e^{-i(\phi_1 - \phi_2)} \right\rangle \qquad (3)$$

where the $\phi_j$ notation corresponding to the partial wave phase modulation induced by the density fluctuations is introduced as following:

$$\phi_j = \frac{\omega^2}{2c^2} \int_0^x \frac{dx_j}{k(x_j)} \frac{\delta n}{n_c}\left(x_j, y\left(y_j, x, x_j, k_j\right)\right)$$

It is natural to assume that the random phases in the exponent been integrals of independent equally distributed random values over long intervals are normally distributed due to the central limit theorem and that they are jointly normally distributed. This assumption allows performing the averaging of the exponential factor in (3) explicitly

$$\left\langle \exp\left[-i(\phi_1 - \phi_2)\right] \right\rangle = \exp\left[-\frac{\left\langle (\phi_1 - \phi_2)^2 \right\rangle}{2}\right].$$

Supposing that the plasma turbulence is weakly statistically inhomogeneous so that its amplitude variation scale length is large compared to the density fluctuations radial correlation length we can simplify the expression for the phase cross-correlation function in the following way

$$\left\langle \phi_i, \phi_j \right\rangle = \frac{1}{4}\frac{\omega^4}{c^4} \int \frac{dx_m}{k(x_m)} \frac{dx_l}{k(x_l)} \left\langle \frac{\delta n(x_m, y'(y_i, x_i, x_m, k_{yi}))}{n_c} \cdot \frac{\delta n(x_l, y'(y_j, x_j, x_l, k_{yj}))}{n_c} \right\rangle =$$

$$= \frac{1}{4}\frac{\omega^4}{c^4} \int \frac{dX}{k^2(X)} \left(\frac{\delta n(X)}{n_c}\right)^2 \int d\Delta\, K\left[\Delta, y'(y_i, x_i, X, k_{yi}) - y'(y_j, x_j, X, k_{yj})\right]$$

where $X = \dfrac{x_m + x_l}{2}, \Delta = x_m - x_l$ and $K(x, y)$ is the two-point cross-correlation function of plasma density fluctuations, which for the sake of simplicity we supposed to be independent of radial coordinate $X$.

The next assumption we need to make is that strong phase modulation regime $<\varphi_i^2>\gg 1$ occurs. In this regime the poloidal correlation length of exponential factor $e^{-i(\varphi_1 - \varphi_2)}$ and more generally of electromagnetic field is much less then plasma density fluctuation poloidal correlation length. Under these circumstances in the domain of high field correlation we may use the following transformation based on the Taylor series expansion of $K(x,y)$:

$$\int d\Delta\, K(\Delta, y'(y_i,x_i,X,k_{yi})-y'(y_j,x_j,X,k_{yj})) = \int d\Delta \iint \frac{d\kappa_x d\kappa_y}{(2\pi)^2}|\delta n|^2_{\kappa_x,\kappa_y}\, e^{i\kappa_x\Delta+i\kappa_y(y'(y_i,x_i,X,k_{yi})-y'(y_j,x_j,X,k_{yj}))} =$$

$$= \int \frac{d\kappa_y}{2\pi}\left|\frac{\delta n^2}{n_c}\right|_{0,\kappa_y} e^{i\kappa_y(y'(y_i,x_i,X,k_{yi})-y'(y_j,x_j,X,k_{yj}))} \approx \int \frac{d\kappa_y}{2\pi}|\delta n|^2_{0,\kappa_y}\left(1-\frac{\kappa_y^{\ 2}}{2}\left(y'(y_i,x_i,X,k_{yi})-y'(y_j,x_j,X,k_{yj})\right)^2\right)$$

The first term of the series expansion proportional to $\kappa_y$ turns to zero after integration due to assumed symmetry of the wavenumber spectrum of the turbulence. So the averaged factor in the Eq. (3) takes the following form:

$$\exp\left[-\frac{\left\langle(\phi_1-\phi_2)^2\right\rangle}{2}\right] = \exp\left[-\frac{\left\langle\phi_1^{\ 2}\right\rangle+\left\langle\phi_2^{\ 2}\right\rangle-2\left\langle\phi_1\cdot\phi_2\right\rangle}{2}\right] =$$

$$= \exp\left[-\frac{1}{8}\frac{\omega^4}{c^4}\int\frac{dX}{k^2(X)}\left(\frac{\delta n(X)}{n_c}\right)^2\left(y'(y_1,x_1,X,k_{y1})-y'(y_2,x_2,X,k_{y2})\right)^2\int\frac{d\kappa_y}{2\pi}\kappa_y^{\ 2}|\delta n|^2_{0,\kappa_y}\right]$$

Using these relations we finally obtain the following expression for the electric field CCF

$$\left\langle E(x,y_1)E^*(x,y_2)\right\rangle = \iint \frac{dk_y dq_y A(k_y)A(q_y)}{(2\pi)^2 k(x)} e^{ik_y y_1 - iq_y y_2 - \frac{i}{2}k_y^2 d^2(x) + \frac{i}{2}q_y^2 d^2(x) - \int\limits_0^x \left((y_1-y_2)-(k_y-q_y)l^2(x,x')\right)^2 D(x')dx'} \qquad (4)$$

where

$$D(x) \triangleq \frac{\omega^4}{c^4}\frac{1}{16\pi k^2(x)}\left(\frac{\delta n(X)}{n_c}\right)^2\int\limits_{-\infty}^{\infty}|\delta n|^2_{0,\kappa_y}\,\kappa_y^2 d\kappa_y$$

The CCF poloidal spectrum averaged across the beam can be explicitly evaluated as

$$\frac{\left|a_{k_y}\right|^2}{k(x)} = \int\left(\int\left\langle E(x,y_1)E^*(x,y_2)\right\rangle d\frac{y_1+y_2}{2}\right)e^{-ik_y(y_1-y_2)}d(y_1-y_2) =$$

$$= \iint d(y_1-y_2)\frac{dq_y}{2\pi\, k(x)}A^2(q_y)e^{iq_y(y_1-y_2)-(y_1-y_2)^2\int\limits_0^x D(x')dx'}e^{-ik_y(y_1-y_2)}$$

It is easy to show that the introduced spectrum $\left|a_{k_y}\right|^2$ satisfies the diffusion equation

$$\frac{\partial\left|a_{k_y}\right|^2}{\partial x} = D(x)\frac{\partial^2\left|a_{k_y}\right|^2}{\partial k_y^{\ 2}} \qquad (5)$$

In the case of Gaussian incident microwave beam $E(0,y)\sim e^{-\frac{y^2}{\delta^2}}$ the solution to (5) is provided by the following averaged CCF poloidal spectrum:

$$\left|a_{k_y}\right|^2 \sim \exp\left[-\frac{k_y^{\ 2}}{\frac{2}{\delta^2}+4\int_0^x D(x')dx'}\right]$$

possessing the angular width given by

$$\sigma^2 = \frac{2}{\delta^2} + 4\int_0^x D(x')dx' \qquad (6)$$

In the case of wave propagation in homogeneous plasma in which statistically homogeneous turbulence is excited the factor $D$ standing for diffusion coefficient is independent of the radial coordinate $x$. In this case the angular width of the beam squared growth linearly with the length of the trajectory x. This kind of dependence is typical for the diffusive phenomena and is characteristic of the photon angular diffusion due to the small-angle-scattering off the long scale density fluctuations. The same broadening law and appropriate diffusivity equation were obtained in the frame of weak turbulence theory (see Section 2).

The microwave beam spatial width dependence on the trajectory length can be derived using expression (4) at $y_1 = y_2$. The integral evaluation in the case of Gaussian beam, $E(0,y) = \varepsilon_0\, e^{-\frac{y^2}{\delta^2}}$, results in the expression obeying the condition of the beam energy flux conservation

$$< E(x,y)E^*(x,y) >= \frac{\varepsilon_0^2}{\sqrt{2}} \frac{k(0)}{k(x)} \frac{\delta}{w(x)} \exp\left[ -\frac{y^2}{w^2(x)} \right],$$

where the beam spatial width is given by

$$w^2 = 4\left( \frac{\delta^2}{8} + \frac{d^4(x)}{2\delta^2} + \int_0^x D(x')l^4(x,x')dx' \right). \qquad (7)$$

The two first terms of the right side of Eq. (7) if no turbulent fluctuations in plasma ($D = 0$) describe diffraction expansion of the beam. In the case of homogeneous background plasma and constant fluctuation level Eq. (7) takes the form

$$w^2 = \frac{\delta^2}{2} + 2\frac{x^2}{\delta^2 k^2} + \frac{1}{12\pi k^4} \frac{\omega^4}{c^4} \frac{\delta n^2}{n_c^2} \left( \int_{-\infty}^{\infty} |\delta n|^2_{0,\kappa_y} \kappa^2 d\kappa \right) x^3$$

In such conditions the beam width expansion follows $x^3$ dependence, and blows up very quickly. It also should be noted that if the density fluctuations exist only in a turbulent layer contribution to the third term in Eq. (7) is nonzero only in the perturbed area.

As it seen from Eq. (6), (7), the spatial and angular beam widths are related by the equation

$$\frac{d}{dx} k(x) \frac{dw^2(x)}{dx} = \frac{2\sigma^2(x)}{k(x)} \qquad (8)$$

In the case the turbulence is localised at the plasma edge at $0 < x < x_t$ using this expression the beam width evolution in the internal discharge region can be obtained in terms of its parameters in the turbulent zone, and at the turbulence zone exit inside the plasma $x = x_t$

$$w^2(x) = w^2(x_t) + 2l^2(x,x_t)\int_0^{x_t} \sigma^2(x') \frac{dx'}{k(x')} + \sigma^2(x_t) l^4(x,x_t)$$

Taking into account the relation $\Delta y = k_y l^2(x, x_t)$ between the poloidal shift of the ray trajectory $\Delta y$ and the corresponding poloidal wavenumber $k_y$ and keeping in mind that in our case $k_y = \sigma(x_t)$ we get

$$w^2(x) = w^2(x_t) + 2\Delta y \int_0^{x_t} \frac{\sigma^2(x')dx'}{\sigma(x_t)k(x')} + (\Delta y)^2 \qquad (9)$$

The later expression derived under the paraxial approximation assumptions can be easily generalized for an arbitrary case of non-slab geometry and beam possessing large angular width. Far from the edge region (at $x \gg x_t$) the spatial beam width is determined by its angular broadening so that the following asymptotic relation is valid $w(x) \simeq \Delta y$. This means that the beam width increases faster than the natural Gaussian beam spatial evolution (see fig. 4e) even if there is no more density fluctuations.

## 2. Beam angular broadening in the frame of weak turbulence theory approach

The approach of weak turbulence theory [12] was recently applied for examination of diffusive regime of wave propagation appearing due to Bragg scattering in 1D geometry [15]. Here we study the 2D geometrical effects of small angle scattering on wave propagation in magnetized plasmas with this method.

We represent density perturbations as a superposition of harmonics in radial and poloidal directions (resp. $x$ and $y$ directions).

$$\frac{\delta n}{n_c} = \sum_{\kappa_x,\kappa_y} \left(\frac{\delta n}{n_c}\right)_{\kappa_x,\kappa_y} e^{i\kappa_x x + i\kappa_y y}$$

where $\kappa_x$ and $\kappa_y$ are the wavenumbers associated to the density fluctuations.

A solution of Eq. (1) can be expressed in the form of Fourier transform

$$E(x,y) = \frac{1}{2\pi}\int E_{k_y}(x) e^{ik_y y} dk_y, \quad E_{k_y}(x) = \frac{a_{k_y}(x) e^{i\int^x \sqrt{k(x')^2 - k_y^2}\,dx'}}{\sqrt[4]{k(x)^2 - k_y^2}} \tag{10}$$

where $a_{k_y}(x)$ stand for amplitudes of incident waves, which can change weakly because of interaction with long wave turbulence. We assume that long wave harmonics are dominant in a drift turbulence spectrum so the scattering is primarily small angle and cut-off point is far so the reflected waves can be omitted.

Acting in a way of weak turbulence theory exactly like in the case of 1D geometry [15] we obtain an integro-differential equation

$$\begin{aligned} \frac{\partial |a_{k_y}|^2}{\partial x} &= \frac{\omega^4}{c^4}\frac{\delta n^2}{n_c^2}\int \frac{d\kappa_y}{8\pi} |a_{k_y-\kappa_y}|^2 |\delta n|^2_{\kappa_x,\kappa_y} \frac{1}{\sqrt{k(x)^2 - k_y^2}\sqrt{k(x)^2 - (k_y-\kappa_y)^2}} - \\ &- \frac{\omega^4}{c^4}\frac{\delta n^2}{n_c^2}\int \frac{d\kappa_y}{8\pi} |a_{k_y}|^2 |\delta n|^2_{\kappa_x,\kappa_y} \frac{1}{\sqrt{k(x)^2 - k_y^2}\sqrt{k(x)^2 - (k_y-\kappa_y)^2}} \end{aligned} \tag{11}$$

The first term on the right hand sides describes increasing of wave energy and the second term describes decreasing of energy due to scattering. The integration over $\kappa_y$ in this case where reflected wave amplitude is negligible is performed along the semicircle performed on the Fig. 1.

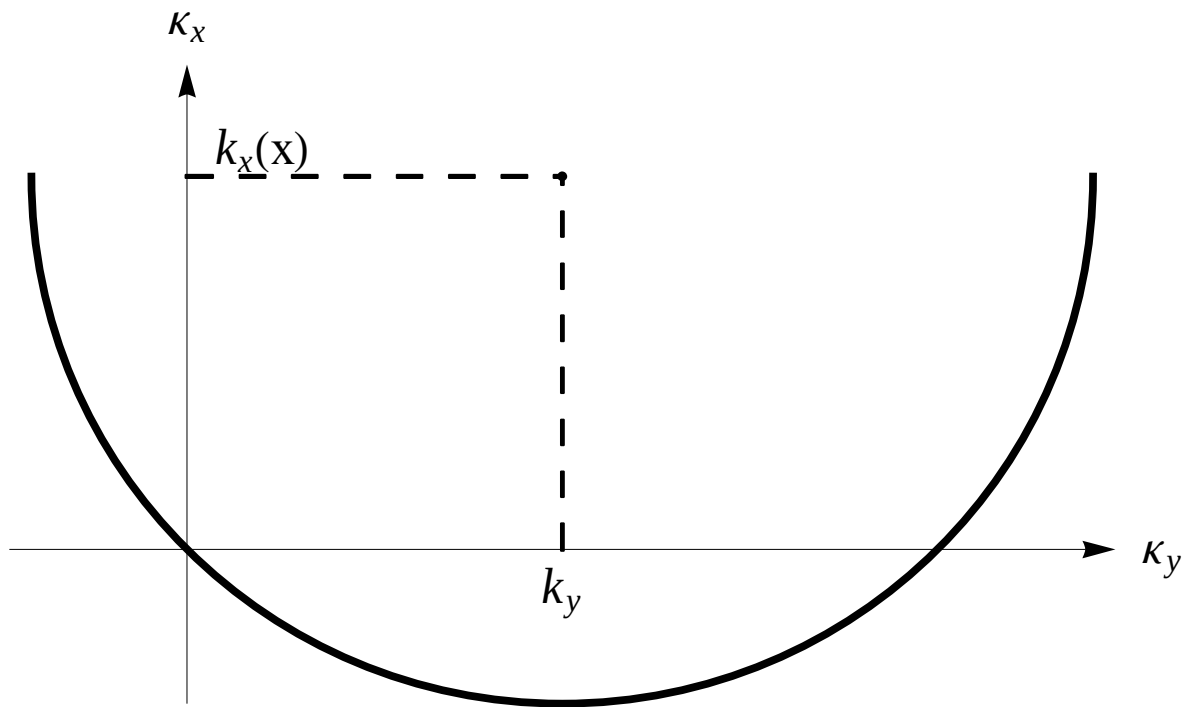

Fig. 1 Wavenumber diagram illustrating the different parts of k-spectrum of the density fluctuations involved in the interaction.

Assuming that the turbulence correlation length is much larger than the beam wave length we may expand integrands in series in terms of $\kappa_y$ near the point $\boldsymbol{\kappa}=0$ and change the integration line there from the circle to the tangent line. Finally, supposing symmetry of fluctuation spectrum we get from Eq. (11) the following equation:

$$\frac{\partial \left|a_{k_y}\right|^2}{\partial x} = \left(\sqrt{1-\frac{k_y^{\,2}}{k(x)^2}}\right)^3 \frac{\partial}{\partial k_y} \frac{D(x)}{1-k_y^{\,2}/k(x)^2} \frac{\partial \left|a_{k_y}\right|^2}{\partial k_y} \qquad (12)$$

where $D(x)$ is the coefficient introduced in the Section 2.

To obtain useful expression, we consider wide Gaussian beam having a small angular width which is characterized by the typical value $\left|k_y\right| \ll |k|$, under this assumption it is possible to simplify the equation providing the beam diagram and obtain it in the form

$$\frac{\partial \left|a_{ik_y}\right|^2}{\partial x} = D(x) \frac{\partial^2 \left|a_{ik_y}\right|^2}{\partial k_y^{\,2}} \qquad (13)$$

which describes diffusive broadening of the incident beam in space of transverse wave numbers $k_y$ and coincides with the Eq. (5).

To illustrate in details the previous case, we look at low background density level $\delta n_o$ and homogeneously distributed cylindrical blobs having Gaussian density profile $\delta n = \delta n_0 e^{-4\frac{x^2+y^2}{L_b^2}}$ ($L_b$ is perpendicular filament size). In this case the angular diffusion coefficient takes the form

$$D = \frac{\sqrt{2\pi}}{4} \frac{\delta n^2}{n_c^{\,2}} \frac{1}{L_b},$$

This expression differs by a factor $\frac{\pi}{8}$ from the analytical estimation of angular diffusion coefficient obtained in [9] based on idea of ray diffusion. The direct comparison of this diffusion coefficient to the result obtained in [10] is unfortunately not possible due to the questionable spherically symmetric blob model utilized in this paper.

It should been mentioned that the analytical results [9, 10] correspond to oversimplified theories, which ignore the plasma and turbulence inhomogeneities and do not deal with the beam spatial width evolution. On contrary, the formulas (6) – (9) derived in this paper permit to evaluate the beam angular and spatial characteristic during its propagation in the inhomogeneous plasma through statistically inhomogeneous turbulence even if the crossed density fluctuations at the plasma edge have high amplitudes. Now to confirm that such models are relevant in realistic cases a comparison to 2D full-wave simulation is done.

## 3. Comparison to the results of 2D full-wave simulations

Comparisons between analytical predictions and averaged results provided by a 2D full-wave code solving Maxwell's equations coupled with J-solver [16] is performed in three cases. An average procedure is needed to recover Gaussian beam shape and its requires a lot of numerical resources to compute the mean value of the wave intensity resulting of 20 independent moving turbulent plasma matrices having the same wavenumber spectrum and the same density fluctuations profile averaged over 8 $10^5$ time steps. The typical computation time on a single processor is around one week per case.

The first two model cases deal with a linear density profile and turbulence, which is statistically homogeneous. For these studies we use different values for density fluctuations (RMS) levels 0.5%, 1%, 1.75%, 2.5%, 3.7%, 5% of critical density at the launched frequency. The isotropic density fluctuations perpendicular wavenumber spectrums are supposed Gaussian with correlations lengths equal to $5\lambda_0$ in the first case and $\lambda_0$ in the second case. The first chosen spectrum contains wavenumbers able only to drive forward scattering whereas Bragg backscattering is also possible in the second case. The input k-spectra as well as density profile are shown in Fig. 2. The microwave frequency considered here is equal to 105 GHz.

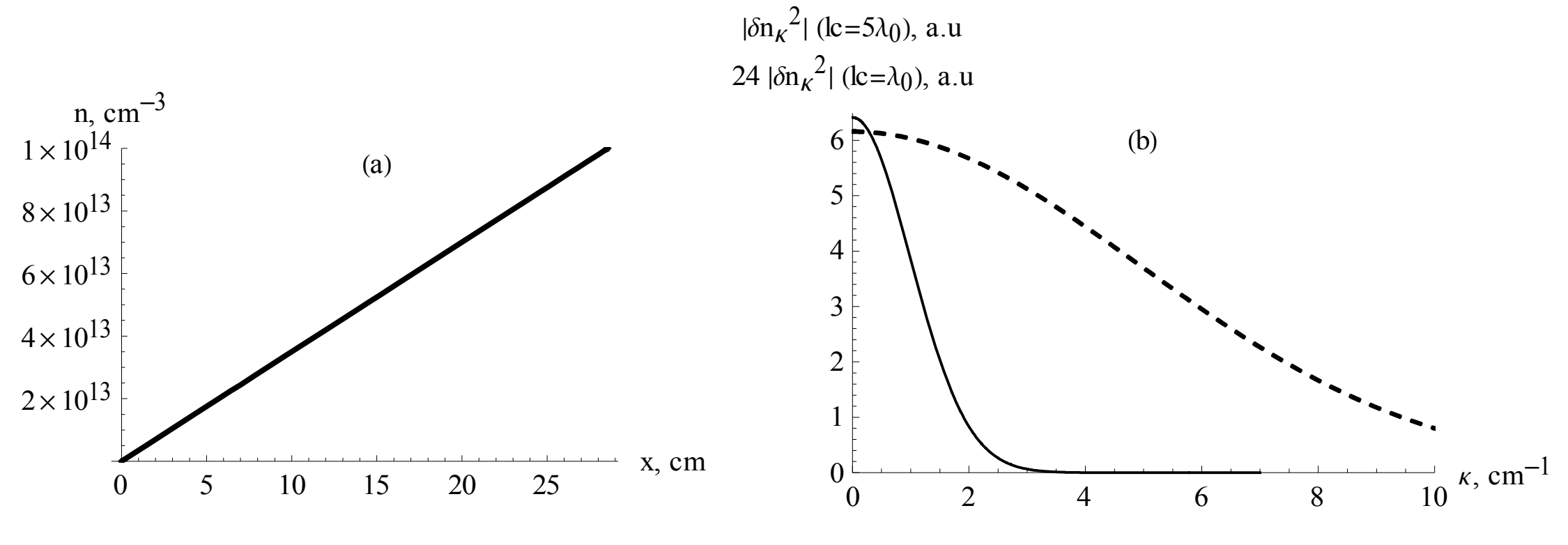

Fig. 2 (a) – Density profile in the first two cases; (b) – Density fluctuation wavenumber spectrums with correlations lengths equal 5λ0 (solid line) and density fluctuation wavenumber spectrums with correlations lengths equal λ0 multiplied by 24 (broken line)

The comparison of theory and simulations for different fluctuation levels in the first case are presented in the Fig. 3a. There is a good agreement between analytical predictions for the beam spatial broadening and numerical modelling results in this case although angular beam broadening may be significant (see Fig. 3b). For the highest level of turbulence the perturbed beam width is already comparable to the computation domain, therefore the beam width saturation there comes from the perfect matching layer used at the mesh border, which remove part of the multi-scattering effects. At the same time, it is shown clearly that ray tracing is irrelevant to compute such effects.

The phase perturbations RMS squared in this case grows linearly with the wave trajectory length (Fig. 3c). It should be mentioned that in this case agreement is good enough both for $\langle \phi \cdot \phi \rangle \gg 1$ and $\langle \phi \cdot \phi \rangle < 1$ .

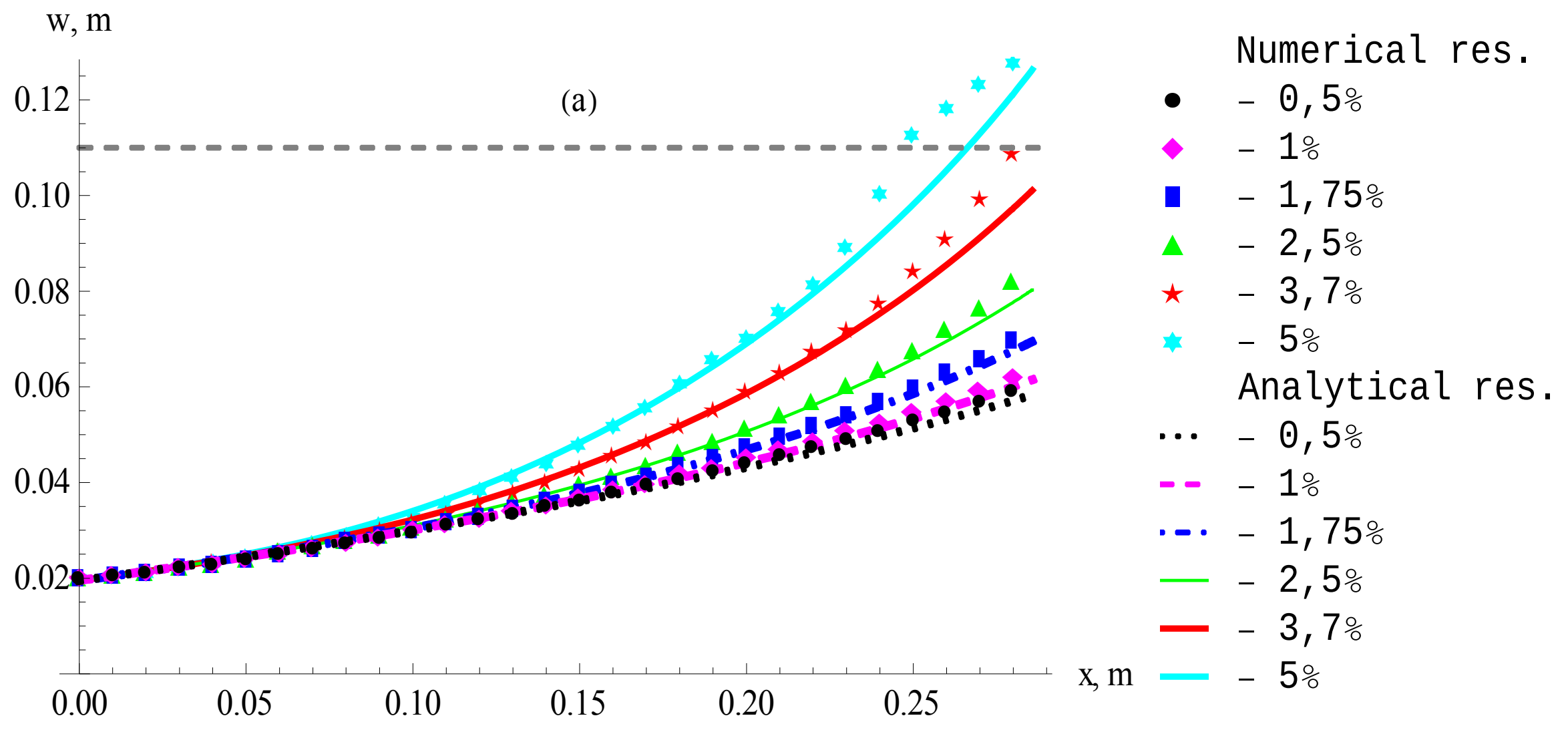

Fig. 3a – The theoretical predictions on the beam width (line) versus numerical results (points) in the case of linear density profile and statistical homogeneous turbulence possessing a Gaussian k-spectrum and correlation length of $5\lambda_o$. The broken line corresponds to the zone where the border effects are important.

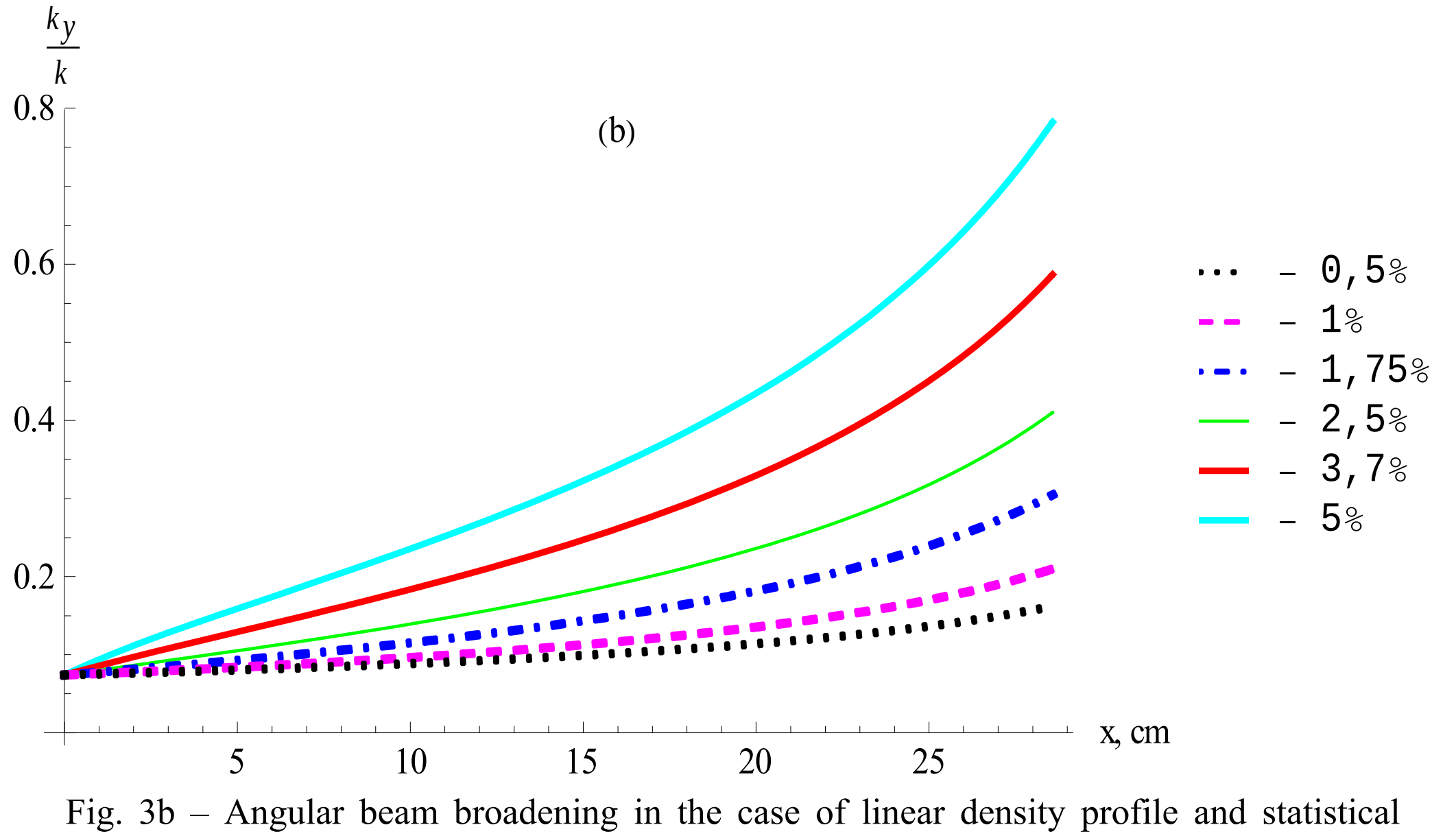

Fig. 3b – Angular beam broadening in the case of linear density profile and statistical homogeneous turbulence possessing a Gaussian k-spectrum and correlation length of $5\lambda_o$.

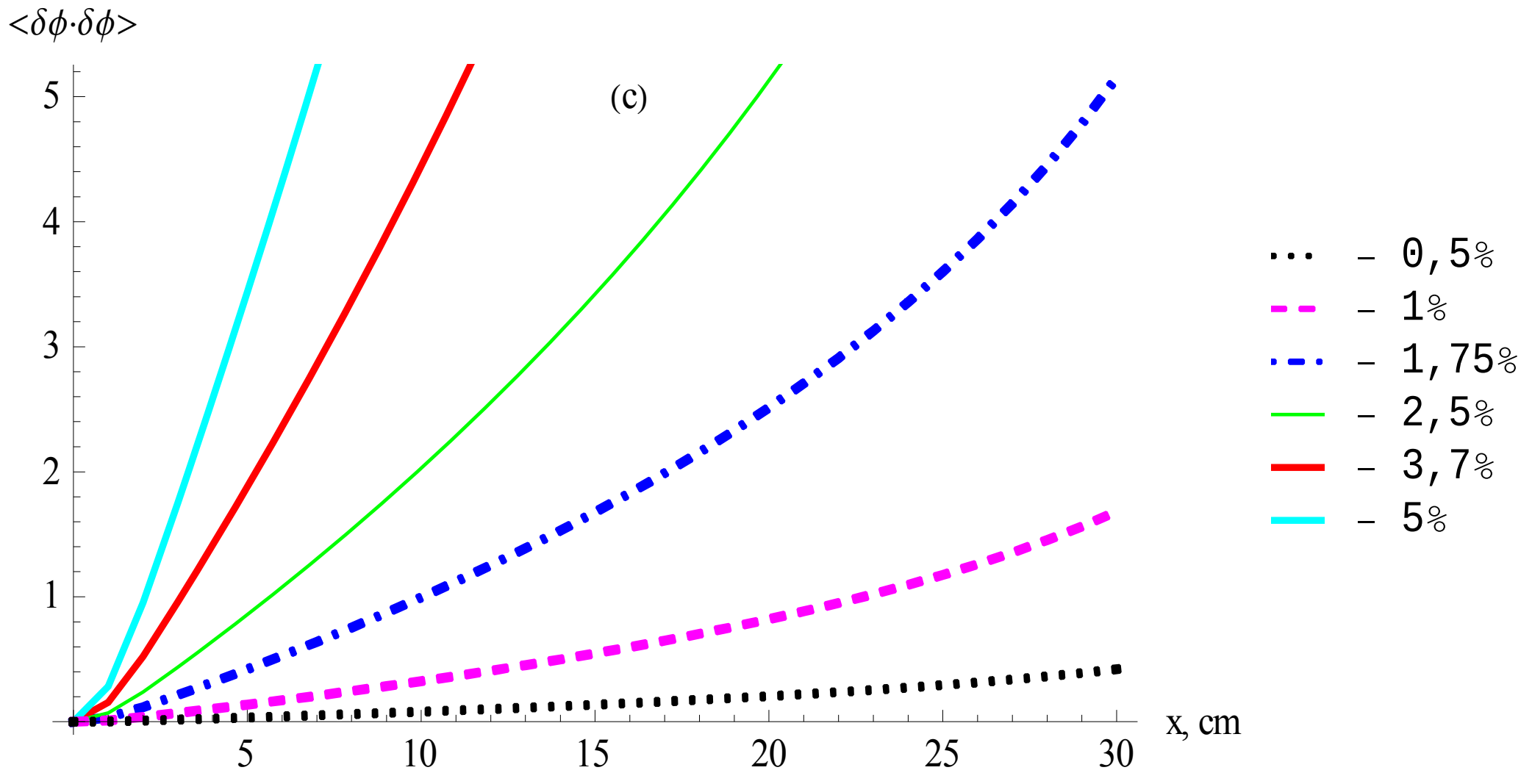

Fig. 3c – The phase RMS squared level in the case of linear density profile and statistical homogeneous turbulence possessing a Gaussian k-spectrum and correlation length of $5\lambda_o$.

Results of comparison in the second case where the WKB approximation is at the applicability border are shown on the Fig. 4a. There is a critical width marked by broken line for which border effects become more important due to a larger widening associated to the change of the k-spectrum shape. On the wave trajectory from boundary to these points angular beam width $k_y/k$ grows up to 0.8 (see Fig. 4b) nevertheless results are in reasonable agreement on this part of wave path. The phase perturbations RMS squared is presented on the Fig. 4c. As it is seen the strong phase perturbation regime is achieved in this case only for large turbulence levels (5%, 3.7% and 2.5%) demonstrating reasonable agreement of analytics and numeric. At smaller levels the difference between broadening provided by the two approaches is more pronounced, nevertheless agreement there is still reasonable.

The third considered case uses the ITER-like plasma parameters, including all the possible inhomogeneities described by the presented models, and those associated to electron cyclotron heating or current drive. The frequency was taken equal to 170 GHz. Density fluctuations RMS and density profiles and the turbulence perpendicular wavenumber spectrum used in the analysis are presented on the Fig. 5a, 5b. The spectrum density, which corresponds to a rescaled wavenumber spectrum coming from experiment [7], is defined by the expression

$$\left|\delta n\right|^2_{\kappa_x,\kappa_y} = \frac{1}{\left(1+0.5\sqrt{\kappa_x^{\ 2}+\kappa_y^{\ 2}}+\left(0.125\sqrt{\kappa_x^{\ 2}+\kappa_y^{\ 2}}\right)^3\right)^2}$$

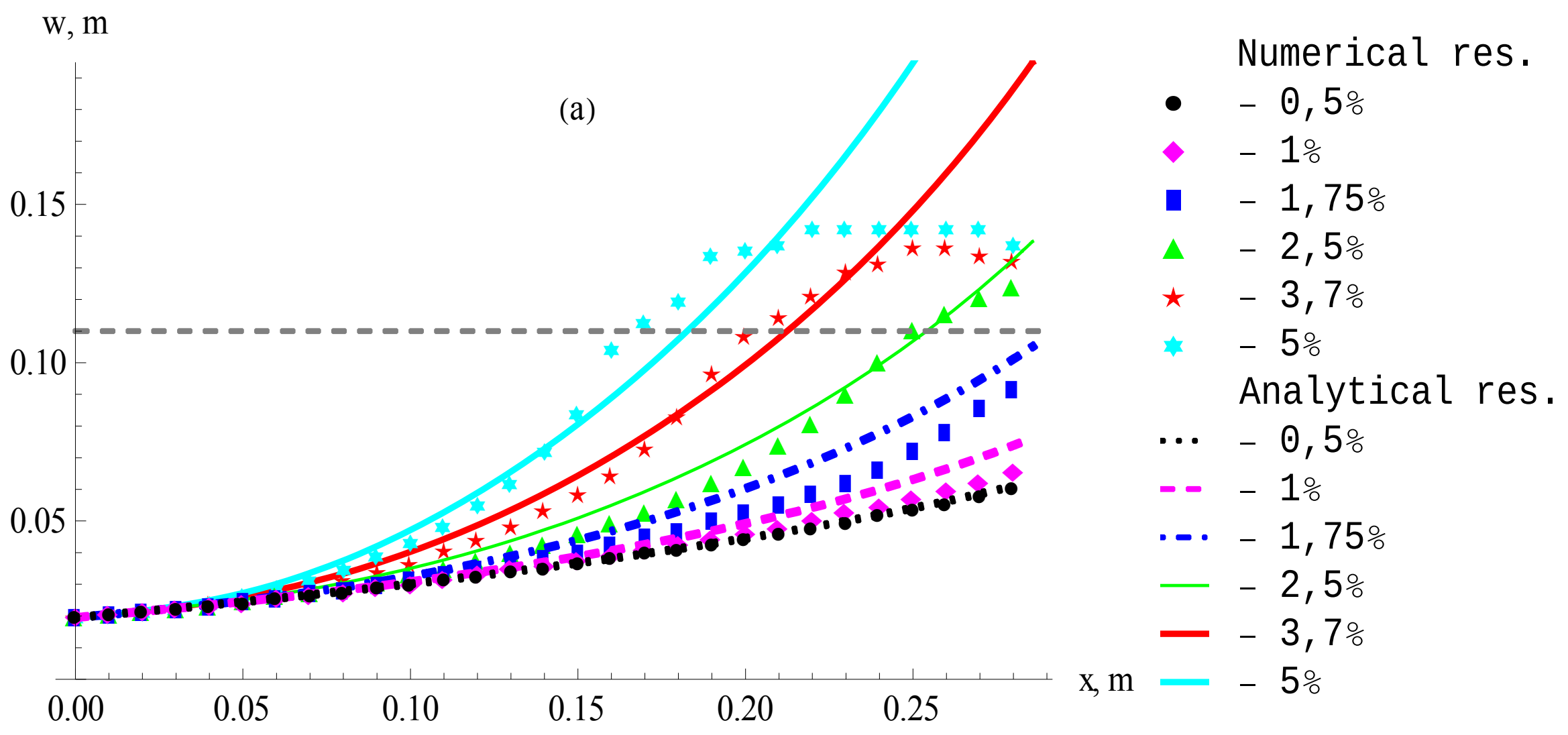

Fig. 4a – The theoretical predictions on the beam width (line) versus numerical results (points) in the case of linear density profile and statistical homogeneous turbulence possessing a Gaussian k-spectrum (correlation length $\lambda_o$).

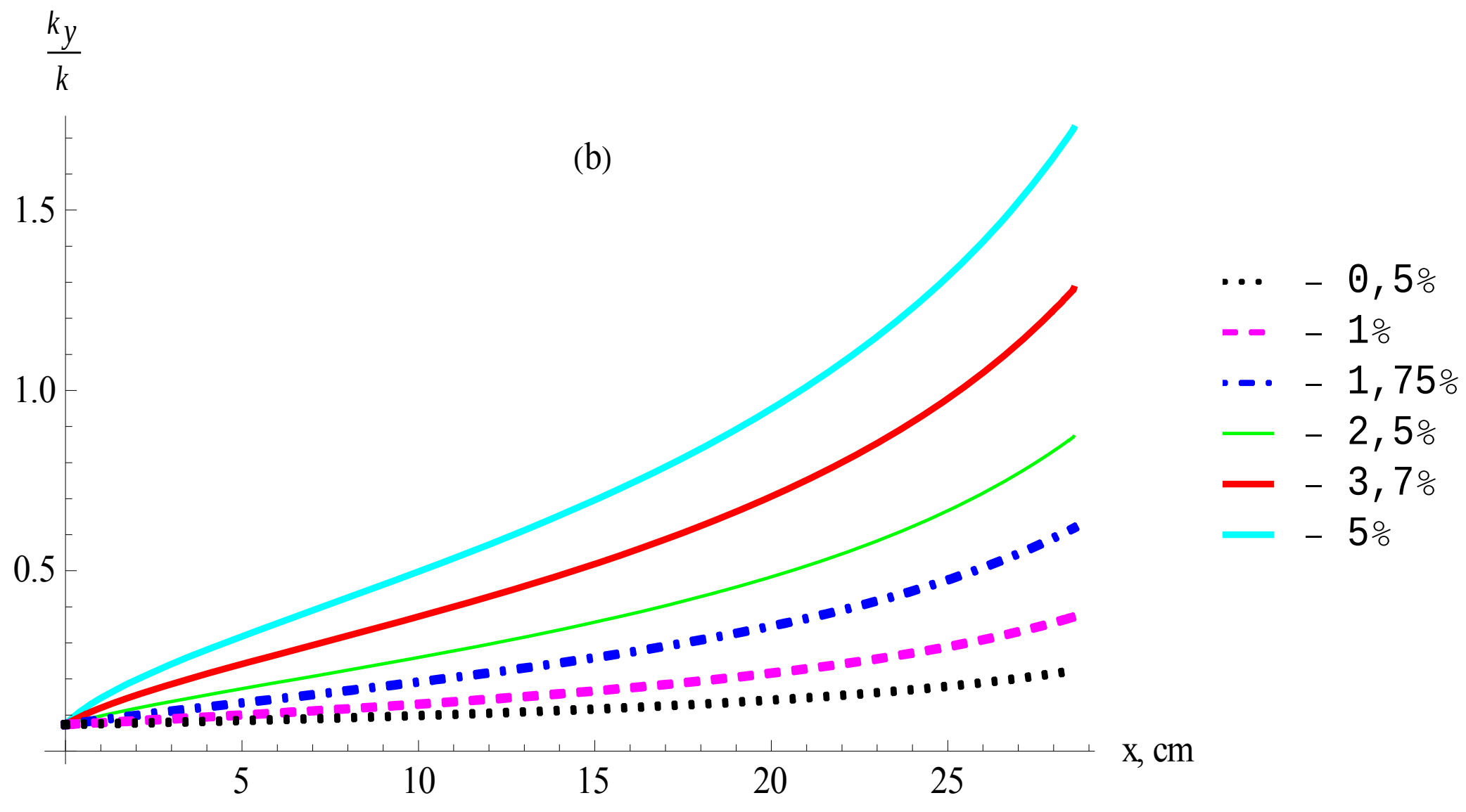

Fig. 4b – Angular beam broadening in the case of linear density profile and statistical homogeneous turbulence possessing a Gaussian k-spectrum (correlation length $\lambda_o$).

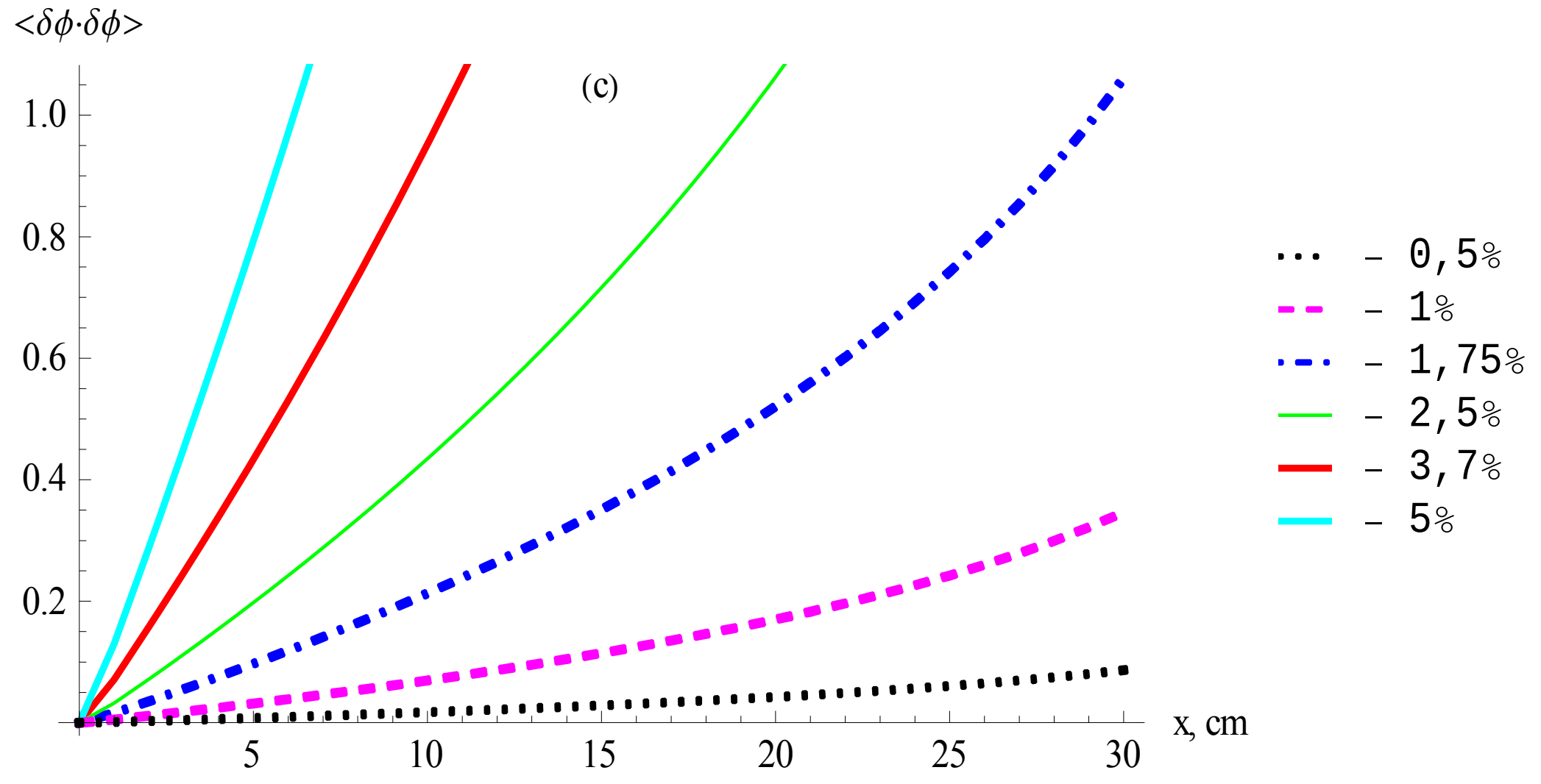

Fig. 4c – The phase RMS squared level in the case of linear density profile and statistical homogeneous turbulence possessing a Gaussian k-spectrum (correlation length $\lambda_o$).

As it is shown in Fig. 5d, the phase perturbation RMS squared in this case grows quickly and exceeds unity near the maximum of the density fluctuations. It seems that small divergence between analytical and numerical results which can be seen in the Fig. 5e is arising from mistakes of the paraxial approximation on the long trajectory and can be eliminate using correct ray trajectories launched into the plasma accounting for the angular broadening in the edge region obtained using paraxial approximation. The corresponding microwave beam width according to (9) is given by expression

$$w(x) \simeq w(x_t) + \frac{\sigma(x_t)}{\sqrt{2}} \int_{x_t}^{x} \frac{dx'}{\sqrt{k^2(x') - \sigma^2(x_t)}}, \qquad (14)$$

As it is seen in Fig.5e, accounting for the non-paraxial effects improves the agreement of analytical predictions and numerical results. It should be noted that this method can be helpful also for prediction of the beam width on the long trajectory far from the edge turbulent layer, where the slab geometry is no longer applicable.

Based on the results of analytical and numerical treatment performed for the ITER case one may state that in spite of the fact the spatial broadening of the O-mode beam in the turbulent layer is not that large, the substantial angular broadening arisen there leads to the further spatial broadening during the propagation along the ray trajectory, which can be very large in the reactor. This effect is not pronounced in the present day devices due to their smaller scale.

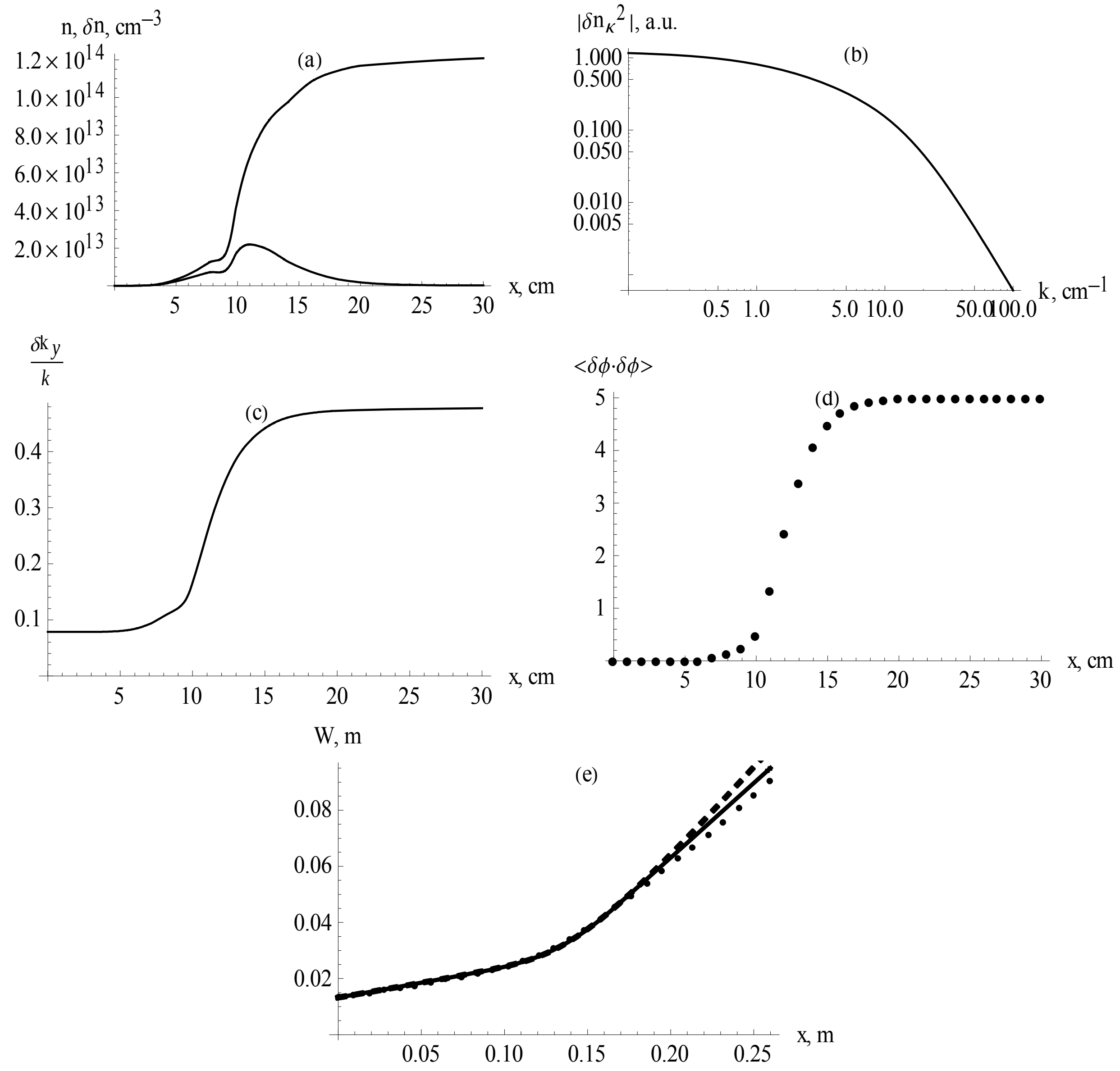

Fig. 5 (a) – Density and fluctuation profiles in the ITER-like case; (b) - density fluctuation wavenumber spectrum; (c) – angular broadening; (d) – the phase RMS squared level; (e) – The paraxial approximation predictions on the beam width (broken line); correct ray tracing based on the angular broadening in the edge region obtained using paraxial approximation (solid line) versus numerical modeling results (dots).

## 4. Conclusions

The analytical evaluation of the impact of the turbulence present at the plasma edge on angular and spatial broadening of the Gaussian beam used for O-mode Electron Cyclotron Resonant Heating have been performed. The eikonal perturbation method has been successfully applied in this paper to describe analytically the broadening of ECRH O-mode beams in edge inhomogeneous and turbulent plasma layer. Simple analytical expressions for diffusion-like angular and spatial beam width variation have been obtained. Diffusion law describing angular broadening in turbulent plasma has been derived also with the weak turbulence theory approach. The developed approach permits to treat the cases of strongly inhomogeneous plasma and turbulence which is typical for edge of fusion machines. The results obtained were compared to full-wave simulation showing a close behaviour in the studied range of turbulence level.
The strong spatial broadening of microwave beam predicted by the two approaches at realistic turbulent edge parameters for ITER ECRH should be accounted for when planning the neoclassical tearing mode control experiment. One can notice the developed analytical method can be as well used to study the reduction of the microwave diagnostic performance induced by turbulence in the edge plasma and on the probing trajectory.
It also should be noted that if only edge plasma zone is perturbed it's possible to use the angular and spatial width defined with the derived expressions as an input for ray-tracing and beam-tracing codes after the turbulence zone to find the beam parameters variation deep in plasma along the ray trajectories evaluated in more complicated geometry.
It should be also noted that in reactor size fusion machines one can expect substantial broadening of X-mode heating and diagnostic beams, as well. The eikonal perturbation method is applicable in these situations as well. The related results will be presented in the forthcoming paper.